\documentstyle[preprint,prd,aps,amssymb,epsfig]{revtex}
\tighten

\def\I{{\rm i}}
\def\R{\mbox{Re }}

\begin{document}
\draft
\preprint{MZ-TH/02-02}
\title{Sweeping the Space of\\ Admissible Quark Mass Matrices}
\author{Silke Falk\cite{SF}, Rainer H\H au\ss ling\cite{RH}, and Florian
  Scheck\cite{FS}} 
\address{Institut f\H ur Physik\\Johannes Gutenberg-Universit\H at\\
D-55099 Mainz (Germany)}
\date{\today}
\maketitle
\begin{abstract}
  We propose a new and efficient method of reconstructing quark mass matrices
  from their eigenvalues and a complete set of mixing observables. By a
  combination of the principle of NNI bases which are known to cover the general
  case, and of the polar decomposition theorem that allows to convert arbitrary
  nonsingular matrices to triangular form, we achieve a parameterization where
  the remaining freedom is reduced to one complex parameter. While this
  parameter runs through the domain bounded by the circle with radius
  $R=\sqrt{(m_t^2-m_u^2)/(m_t^2-m_c^2)}$ around the origin in the complex plane
  one sweeps the space of all mass matrices compatible with the given set of
  data. 
\end{abstract}

\vfill
\section{Introduction}
The four three-generation mass sectors of quarks and leptons belong to
the deepest enigmas of the standard model of strong and electroweak
interactions. While there is a great amount of experimental
information of steadily increasing accuracy, theoretical models of
mass matrices and mixing matrices are scarce. In this unsatisfactory
situation it seems of utmost importance to parameterize the available
data in such a way that possible textures in the mass matrices of a
given charge sector become visible in an unambiguous manner. For
instance, in the case of quarks, there are 10 data, viz. the masses of
the three charge~$+2/3$ quarks, the masses of the three
charge~$-1/3$ quarks, and four observables in the
Cabibbo-Kobayashi-Maskawa (CKM) matrix, to be compared to 12
physically significant parameters in the mass matrices $M^{(u)}$ and
$M^{(d)}$. Therefore, reconstructing mass matrices from the data
really amounts to finding an optimal parameterization that exhibits
the remaining two-parameter freedom in a transparent way.

An important step in this direction was taken by Branco et al. who
realized that the class of so-called nearest-neighbour-interaction
(NNI)-bases for chiral states are economical but still completely
general \cite{BLM}, and may hence be used in attempts to reconstruct
the mass matrices from the observed mass eigenvalues and the empirical
mixing matrices.  These authors also gave an explicit procedure for
constructing mass matrices in an NNI basis, for arbitrary mass sectors
of quarks. Unfortunately, their analysis involves solving cubic
equations. Although soluble in principle, these equations are too
cumbersome to solve and do not allow for a practical and efficient
reconstruction.

In this paper we propose a new method of reconstruction that avoids
these shortcomings. We conjecture that this method and the appropriate
parameterization are optimal in the sense of concentrating the
remaining freedom in a single complex parameter whose domain of
variation can be restricted to the interior of a circle in the complex
plane. In particular, we succeed in reconstructing the mass matrices proper, up
to unobservable changes of basis. This goes beyond, say, the work of Harayama
and Okamura \cite{HaO} who express the CKM~matrix in terms of six parameters,
with a two-parameter freedom. The way from their result to the mass matrices
seems involved and not well suited for a practical analysis.

We make use of the polar decomposition theorem for nonsingular
matrices \cite{Mur}
\begin{displaymath}
  M\, =\, T\, W\;\, ,
\end{displaymath}
where $T$ is a lower-triangular matrix and $W$ is a unitary matrix
which, when applying this formula to three generations of chiral quarks,
can be absorbed in the right-chiral fields. If, in addition, we work
in the class of NNI bases in which the $(21)$-element of $T$ is seen to
vanish, we still cover the most general case but get rid of all
redundant quantities. More precisely, if
\begin{displaymath}
  \widehat{H}^{(q)} = M^{(q)}M^{(q)\dagger} = 
  \widehat{T}^{(q)}\widehat{T}^{(q)\dagger}\, ,\qquad q= u,d\, ,
\end{displaymath}
are the ``squared'' hermitean mass matrices and $D^{(q)}=\mbox{diag }
\bigl( m_1^2,m_2^2,m_3^2\bigr)$ with $m_1\equiv m_u$ or $m_d$ etc.
their diagonal forms, then, in any NNI basis,
\begin{displaymath}
  \widehat{H}^{(u)} = U^\dagger D^{(u)} U\, ,\;\;
  \widehat{H}^{(d)} = U^\dagger V_{\rm CKM} D^{(d)} 
  V_{\rm CKM}^\dagger U\, .
\end{displaymath}
The matrix $U$ which is known analytically \cite{HS1}, depends on two
complex parameters, say $a$ and $b$ defined in eq.~(\ref{newvar})
below. These parameters are related through a \emph{quadratic}
equation whose coefficients are elements of the matrix
\begin{displaymath}
  V_{\rm CKM} D^{(d)} V_{\rm CKM}^\dagger\, ,
\end{displaymath}
i.e. of a matrix that is obtained solely from experimental data!
Solving for one or the other of them, say $b=b(a)$, reduces the set of
admissible mass matrices to the expected two-parameter freedom in the variable $a$.
Progress achieved in this way is twofold: On the one hand,
parameterization in terms of, say, $a$ is analytically simple and
transparent. On the other hand, the domain of variation of $a$ and the 
formulae for the elements of $\widehat{H}^{(u)}$ and
$\widehat{H}^{(d)}$ are such that the space of admissible mass
matrices can be studied graphically and numerically, as $a$ sweeps
through all allowed values, in a quantitatively reliable
manner. Although we have not done this yet, one can even follow the
propagation of the error bars of the experimental input, in not too
involved a procedure.

The paper is organised as follows: In sect.~2 we review the choice of
NNI bases and recapitulate the relevance of the polar decomposition
theorem for the problem at stake. Sect.~3 which is the main body of
our work, describes the explicit construction of the matrix $U$ as
well as its parameterization in terms of $a$, $b$, and the squared
masses of the \emph{up-}sector. The NNI condition is encoded in the
quadratic equation (\ref{constr}) below. In sect.~4 we discuss
symmetries helpful in solving that constraint, and consider some
limiting cases in order to illustrate the method. We also propose an
expansion of the solutions in terms of a parameter which, due to the
hierarchy in the quark masses, is numerically very small. The final
sect.~5 gives two examples of how a conjectured texture of mass
matrices can be checked against the data in a simple and transparent
manner, by converting the mass matrices to the (general) form studied
here. It ends with a few conclusions.
\section{General nearest-neighbour bases and triangular matrices}
In the case of three generations the mass matrices $\widehat{M}^{(u)}$
and $\widehat{M}^{(d)}$ in the so called \emph{nearest neighbour
  interaction} (NNI) basis are characterized by the following generic
structure,
\begin{equation}
\label{nnistructure}
  \widehat{M}^{(u)}, \; \widehat{M}^{(d)} \; = \;
  \left( \begin{array}{ccc}
  0 & \star & 0 \\ \star & 0 & \star \\ 0 & \star & \star
  \end{array} \right) \; \; \; .
\end{equation}
Here the ``$\star$''-entries appearing on the r.h.s. of (\ref{nnistructure}) are
arbitrary non-vanishing, complex numbers. The interpretation of this particular
choice of the mass matrices that is usually given by its proponents is the
following: the $(11)$- and $(22)$-elements are put equal to zero, while letting
the $(33)$-element be different from zero, with the idea of describing an
initial, no-interaction situation where two quarks are massless and only one is
massive. Furthermore, only neighbouring generations are allowed to interact, by
assuming nonvanishing $(12)$-, $(21)$-, $(23)$-, and $(32)$-elements, but
vanishing $(13)$- and $(31)$-elements. 

However, it has been known for a long time that this setting, although very
tempting and intuitive at first sight, is ill-defined unless it is supplemented
by further assumptions.  Indeed, Branco, Lavoura, and Mota showed in \cite{BLM}
that any set of admissible mass matrices $\{ M^{(u)}, M^{(d)} \}$ of the
standard model, i.e.  any set of two \emph{completely arbitrary} non-singular $3
\times 3$-matrices, can be transformed to the form given in (\ref{nnistructure})
without the need for any further assumption. In other words, in the framework of
the minimal standard model where only left-chiral fermion fields participate in
charged current weak interactions, the mass matrices in (\ref{nnistructure}) are
still completely general, the specific form (\ref{nnistructure}) reflecting no
more than a specific choice of chiral basis. This fact is mainly due to the
observation that weak interactions of quarks remain unchanged under the
simultaneous transformations
\begin{eqnarray}
\label{masstrans}
  M^{(u)} & \rightarrow & \widehat{M}^{(u)} \; = \; 
  U^\dagger M^{(u)} V_u \; \; \; \mbox{ and} \nonumber \\
  M^{(d)} & \rightarrow & \widehat{M}^{(d)} \; = \; 
  U^\dagger M^{(d)} V_d \; \; \; ,
\end{eqnarray}
of the mass matrices, where $U, V_u$ and $V_d$ are arbitrary unitary $3 \times
3$-matrices: On the one hand, the unitaries $V_u$ and $V_d$ act on the
right-chiral fields $(u_R, c_R, t_R)$ and $(d_R, s_R, b_R)$, respectively, and,
hence, can be absorbed by a redefinition of these unobservable fields, without
loss of generality. On the other hand, the common unitary matrix $U$ in
(\ref{masstrans}) which acts on the left-chiral quark fields, drops out when
calculating the physical Cabibbo-Kobayashi-Maskawa (CKM) matrix
\begin{equation}
\label{vckmdef}
  V_{\rm CKM} \; = \; U^{(u)}_L U^{(d) \dagger}_L \; \; \; .
\end{equation}
In (\ref{vckmdef}) $U^{(u)}_L$ and $U^{(d)}_L$ are the unitary matrices which
diagonalize the ``squared'', hermitean mass matrices
\begin{equation}
  \label{squmat}
  H^{(q)} = M^{(q)} M^{(q)  \dagger}\, ,\qquad q = u, d\, ,
\end{equation}
of the up- and down-sector, respectively, viz.
\begin{eqnarray}
\label{massdiag}
  U^{(u)}_L H^{(u)} U^{(u) \dagger}_L & = & \mbox{diag} 
  (m_u^2, m_c^2, m_t^2) \equiv D^{(u)}\; , \nonumber \\
  U^{(d)}_L H^{(d)} U^{(d) \dagger}_L & = & \mbox{diag}
  (m_d^2, m_s^2, m_b^2) \equiv D^{(d)}\; .
\end{eqnarray}
Thus, as proved in \cite{BLM}, the structure (\ref{nnistructure}) of the mass
matrices corresponds to no more than a special, physically admissible choice of
the electroweak basis. The interpretation sketched above fully rests on this
choice and will no longer be valid in other electroweak bases. Unless this
special choice is singled out by additional arguments that could stem, e.g.,
from physics beyond the standard model, the above interpretation loses its
physical significance because \emph{physics}, of course, \emph{must not depend
  on the choice of basis}.\\
There is an alternative derivation of the same conclusion which, at the same
time, helps us to fix notations for our subsequent calculations. As was shown in
\cite{HS1}, \cite{HS2}, when dealing with questions of reconstructing mass
matrices from the experimental data, i.e. from four independent absolute values
of CKM matrix elements and the six quark masses, a most economic but
nevertheless physically completely general parameterization of the quark mass
matrices is given by \emph{triangular mass matrices},
\begin{equation}
\label{tristructure}
  T^{(u)} , \; T^{(d)} = \left( \begin{array}{ccc}
  \star & 0 & 0 \\ \star & \star & 0 \\ \star & \star & \star
  \end{array} \right) \; \; \; .
\end{equation}
This parameterization results upon exploiting the polar decomposition
theorem for non-singular, but otherwise arbitrary matrices:
\begin{equation}
\label{polardec}
  M^{(q)} \; = \; T^{(q)} W^{(q)} \; ,\quad q = u, d\, , \quad 
    \mbox{with } W^{(q)} \mbox{ unitary}.
\end{equation}
Again, the unitaries $W^{(q)}$, $q = u,d$, can be absorbed by a redefinition
of the right-chiral quark fields, without loss of generality.\\
At the level of the squared mass matrices $\widehat{H}^{(q)}$ the NNI
structure (\ref{nnistructure}) of the mass matrices is equivalent to a vanishing
$(12)$-element (and -- because $\widehat{H}^{(q)}$ is hermitean -- also a
vanishing $(21)$-element):
\begin{equation}
\label{nnicond}
  \widehat{H}^{(q)}_{12} \; = \; 0 \; = 
  \; \widehat{H}^{(q)}_{21} \; \; , \; \; q = u,d \; .
\end{equation}
Because of $\widehat{H}^{(q)} = \widehat{M}^{(q)} \widehat{M}^{(q) \dagger} = 
\widehat{T}^{(q)} \widehat{T}^{(q) \dagger}$,
$q = u,d$, a straightforward calculation shows that within the triangular
parameterization the NNI condition (\ref{nnicond}) reads
\begin{equation}
\label{nnicondtri}
  \widehat{T}^{(q)}_{21} \; = \; 0\; , \qquad q = u, d \; \; \; .
\end{equation}
Mimicking for a moment the (invalid) interpretation mentioned in connection with
the NNI structure (\ref{nnistructure}), eq. (\ref{nnicondtri}) would suggest
that there is \emph{no} direct interaction between the first and the
second generation while, in the basis (\ref{nnistructure}), these evidently do
mix, and, in fact, mix strongly. Direct interactions seem to be present between
the second and third generations as well as between the first and third
generations only, due to the non-vanishing $(32)$- and $(31)$-elements of
$\widehat{T}^{(q)}$, respectively, in contrast to (\ref{nnistructure}) where
seemingly there is no direct coupling between first and third generations.\\
These statements underpin once more that such an interpretation is dependent
on the electroweak basis chosen for the representation of the mass matrices and,
hence, should better be avoided altogether.\\

Although it is unrelated to a specific physical picture of quark masses and
mixings, bases that yield the NNI form of the mass matrices turn
out to be very useful in the process of reconstructing mass matrices from the
observed quark mixings and masses.  Therefore, in what follows we shall make
extensive use of this class of bases. That is to say, we start from
triangular mass matrices whose $(21)$-elements are zero, viz.

\begin{equation}
\label{trinni}
  \widehat{T}^{(u)} = \left( \begin{array}{ccc}
  \hat{\alpha} & 0 & 0 \\ 0 & \hat{\beta} & 0 \\
  \hat{\kappa}_3 e^{\I \hat{\varphi}_3} & 
  \hat{\kappa}_2 e^{\I \hat{\varphi}_2} & \hat{\gamma}
  \end{array} \right) \; \; \; , \; \; \;
  \widehat{T}^{(d)} = \left( \begin{array}{ccc}
  \hat{\alpha}' & 0 & 0 \\ 0 & \hat{\beta}' & 0 \\
  \hat{\kappa}'_3 e^{\I \hat{\varphi}'_3} & 
  \hat{\kappa}'_2 e^{\I \hat{\varphi}'_2} & \hat{\gamma}'
  \end{array} \right)\; .
\end{equation}
Here and in the sequel the hat on the symbols refers to the choice of an
NNI basis.

Possible phases can be absorbed into the right-chiral fields and,
hence, without loss of generality, the matrix elements
$\hat{\alpha}, \hat{\beta}, \hat{\gamma}$ and
$\hat{\alpha}', \hat{\beta}', \hat{\gamma}'$ along the
diagonals may be chosen to be real. In fact, also the phases
$\hat{\varphi}_2$ and $\hat{\varphi}_3$ could be dropped by
making use of the unitary matrix $U$ in (\ref{masstrans}) thereby
rendering $\widehat{T}^{(u)}$ real. In doing so, on the
theoretical side we are left with 12 parameters, 5 from the up-sector
and 7 from the down-sector.  These have to be confronted with 10
experimental data, i.e. 6 quark masses + 4 real observables in the CKM
matrix, leaving a freedom of two parameters.  This is characteristic
for an NNI basis.


\section{An efficient reconstruction procedure}
The authors of \cite{BLM} give a detailed prescription for the
construction of mass matrices in an NNI basis, for arbitrary mass
matrices: The main step consists in solving the eigenvalue problem
\begin{equation}
\label{nnieigen}
  ( H^{(u)} + \kappa H^{(d)} )_{ij} \; u_{j2} \; = \; \lambda \; u_{i2}
\end{equation}
for the matrix $H^{(u)} + \kappa H^{(d)}$, where $\kappa$ denotes an
arbitrary complex number. Note that $\kappa$ reflects the
two-parameter freedom within the class of NNI matrices. Once the
second column $(u_{i2})$ of the unitary matrix $U$, which transforms
the arbitrary squared mass matrices (\ref{squmat}) to NNI form (see
(\ref{masstrans})) is determined according to (\ref{nnieigen}), the
first column $(u_{i1})$ is calculated by means of
\begin{equation}
\label{firstco}
  u_{i1} \; \propto \; \epsilon_{ijk} \; u_{j2}^\star \;
  u_{l2}^\star \; (H^{(u)})_{lk}
  \; \; \; ,
\end{equation}
see \cite{BLM}. Finally, the third column $(u_{i3})$ follows from the unitarity
of $U$. It is easily checked that (\ref{nnieigen}) and (\ref{firstco}) indeed
imply $u_{i1}^\star \; (H^{(q)})_{ij} \; u_{j2}
\; = \; 0$ for $q = u,d$ as required.\\
However, the prescription just outlined is not very well suited when aiming at
an \emph{explicit} construction of all NNI mass matrices.  This is simply due to
the fact that in the interesting case of three generations the eigenvalue
problem (\ref{nnieigen}) leads to a cubic equation. The solutions of this cubic
equation are, of course, known in principle but the corresponding expressions
are rather lengthy and involved and they complicate tremendously subsequent
calculations.  In this paper we propose a different procedure which will be
seen to result in much simpler expressions.\\
We start from the following observation \cite{HS2}: Without loss of generality
we may assume that the squared mass matrix of $u$-quarks is already diagonal,
that is, in other words, that the mixing has been shifted entirely to the
down-sector. Indeed, this is achieved by exploiting once more the freedom
contained in (\ref{masstrans}) by choosing $U = U^{(u) \dagger}_L$. With this
choice we obtain (using (\ref{masstrans}), (\ref{vckmdef}), and
(\ref{massdiag})):
\begin{eqnarray}
\label{startpoint0}
  H^{(u)} & \longmapsto & 
  U^{(u)}_L H^{(u)} U^{(u) \dagger}_L \;
  = \; D^{(u)} \; \; \; \mbox{ and} \\
  H^{(d)} & \longmapsto & 
  U^{(u)}_L H^{(d)} U^{(u) \dagger}_L \;
  = \; U^{(u)}_L U^{(d) \dagger}_L D^{(d)} U^{(d)}_L U^{(u) \dagger}_L 
  \nonumber \\
  & & \hspace*{.95cm} = \; V_{\rm CKM} D^{(d)} V_{\rm CKM}^\dagger \nonumber
\end{eqnarray}
Thus, in what follows, by this redefinition, the experimental input will be
coded in the form
\begin{equation}
\label{startpoint}
  H^{(u)} \; = \; D^{(u)} \; \; \; \mbox{ and } \; \; \;
  H^{(d)} \; = \; V_{\rm CKM} D^{(d)} V_{\rm CKM}^\dagger \; \; \; .
\end{equation}
Note that $H^{(d)}$ is \emph{completely} fixed in terms of the experimental
input, i.e. the three quark masses of the down-sector and four physically
relevant parameters of the CKM matrix. Let us comment on this point in more
detail: Because, in general, normed eigenvectors are only fixed up to arbitrary
phase factors, instead of $U^\dagger = U^{(u)}_L$ we can as well choose
$U^\dagger = P_1 U^{(u)}_L$ in the above reasoning, with $P_1$ a diagonal phase
matrix,
\begin{equation}
\label{phase1}
  P_1 \; = \; \mbox{diag}(e^{\I \phi_1}, e^{\I \phi_2}, e^{\I \phi_3} )
  \; .
\end{equation}
With this choice the second substitution in (\ref{startpoint0}) becomes:
\begin{eqnarray}
\label{arg}
  H^{(d)} & \rightarrow & \widehat{H}^{(d)} \; = \;
  P_1 V_{\rm CKM} \; D^{(d)} \; V_{\rm CKM}^\dagger P_1^\dagger \nonumber \\
  & & \hspace*{.95cm} = \; P_1 V_{\rm CKM} P_2^\dagger \; D^{(d)} \; P_2
  V_{\rm CKM}^\dagger P_1^\dagger
\end{eqnarray}
In the last line, $P_2$ again denotes a diagonal matrix containing only phase
factors,
\begin{equation}
\label{phase2}
  P_2 \; = \; \mbox{diag} (e^{\I \phi_4}, e^{\I \phi_5}, 1)
  \; \; \; ,
\end{equation}
and the diagonal character of $D^{(d)}$ has been used.  Eq.  (\ref{arg}) shows
that we are allowed to choose \emph{any} parameterization for $V_{\rm CKM}$ that
we would like to start with.  The transition to any other parameterization of
$V_{\rm CKM}$ is easily accomplished by means of suitable choices of the phase
matrices $P_1$ and $P_2$.\\
Next, we have to tackle the task of finding the unitary matrices $U$ which
transform the squared mass matrices $H^{(q)}$ to NNI form.  Postponing for a
moment the analysis of the down-sector, the corresponding condition for $U$ in
the up-sector simply reads
\begin{equation}
\label{upcond}
  \widehat{H}^{(u)} \; = \; U^\dagger D^{(u)} U \quad
  \Longleftrightarrow \quad
  D^{(u)} \; = \; U \widehat{H}^{(u)} U^\dagger \; \; \; .
\end{equation}
In other words, at this stage $U$ is given by those unitary matrices that
diagonalise the most general squared mass matrix $\widehat{H}^{(u)} =
\widehat{T}^{(u)} \widehat{T}^{(u) \dagger}$ where $\widehat{T}^{(u)}$ is taken
from (\ref{trinni}) (with $\hat{\varphi}_2 = 0 = \hat{\varphi}_3$,
without loss of generality). This is the first condition for the
matrix $U$.

An analytical expression for the matrix $U$ is obtained by restricting
to the case $\kappa_1 = 0$ the general solution of the problem of
diagonalisation that we had obtained earlier in \cite{HS1}, viz.
\begin{eqnarray}
\label{upre}
  U & = & P \; \cdot \left( \begin{array}{ccc}
  f(m_u)/n_1 & g(m_u)/n_1 & h(m_u)/n_1 \\
  f(m_c)/n_2 & g(m_c)/n_2 & h(m_c)/n_2 \\
  f(m_t)/n_3 & g(m_t)/n_3 & h(m_t)/n_3 \end{array} \right) 
  \;  , 
\end{eqnarray}
where the functions $f(m_i)$, $g(m_i)$, $h(m_i)$ and the
denominators $n_k$ are given by
\begin{eqnarray}
  \begin{array}{r}
  f(m_i) \\ g(m_i) \\ h(m_i) \end{array}
  & \begin{array}{c} = \\ = \\ = \end{array} & \left.
  \begin{array}{l}
   - \hat{\alpha} \hat{\kappa}_3 (\hat{\beta}^2 - m_i^2) \\
   - \hat{\beta} \hat{\kappa}_2 (\hat{\alpha}^2 - m_i^2) \\
  (\hat{\alpha}^2 - m_i^2)(\hat{\beta}^2 - m_i^2)
  \end{array} \right\} \; (i = u,c,t) \nonumber \\
  \begin{array}{r}
  n_1^2 \\ n_2^2 \\ n_3^2 \end{array}
  & \begin{array}{c} = \\ = \\ = \end{array} &
  \begin{array}{l}
  (\hat{\alpha}^2 - m_u^2)(\hat{\beta}^2 - m_u^2)
  (m_t^2 - m_u^2)(m_c^2 - m_u^2) \\
  (m_c^2 - \hat{\alpha}^2)(\hat{\beta}^2 - m_c^2)
  (m_t^2 - m_c^2)(m_c^2 - m_u^2) \\
  (m_t^2 - \hat{\alpha}^2)(m_t^2 - \hat{\beta}^2)
  (m_t^2 - m_c^2)(m_t^2 - m_u^2) \; \; \; . \end{array} \nonumber
\end{eqnarray}
The diagonal phase matrix $P$,
\begin{equation}
\label{phase}
  P \; = \; \mbox{diag} (- 1, e^{\I \psi_2}, e^{\I \psi_3} ) \; \; \; ,
\end{equation}
represents the freedom of multiplying each eigenvector of $\widehat{H}^{(u)}$
with an arbitrary phase factor\footnote{The choice of the first phase, $\exp
  \{\I\pi\}$, is made for the sake of convenience.}. As we are aiming at
\emph{all} NNI mass matrices this freedom must be taken into account. This will
become clear also in a moment when we count the
degrees of freedom explicitly.\\
Furthermore, from the comparison of the characteristic polynomials of
$\widehat{T}^{(u)} \widehat{T}^{(u) \dagger}$ and of $D^{(u)}$ the parameters
$\hat{\kappa}_2$ and $\hat{\kappa}_3$ are fixed in terms of $\hat{\alpha}$,
$\hat{\beta}$ and the squared quark masses of the up-sector by means of the
relations:
\begin{eqnarray}
\label{char1}
  m_u^2 m_c^2 m_t^2 & = & \hat{\alpha}^2 \hat{\beta}^2
  \hat{\gamma}^2 \\ 
\label{char2}
  m_u^2 m_c^2 + m_u^2 m_t^2 + m_c^2 m_t^2 & = &
  \hat{\alpha}^2 \hat{\beta}^2 + \hat{\beta}^2 \hat{\gamma}^2 +
  \hat{\gamma}^2 \hat{\alpha}^2 + \hat{\alpha}^2
  \hat{\kappa}_2^2 + 
  \hat{\beta}^2 \hat{\kappa}_3^2 \\
\label{char3}
  m_u^2 + m_c^2 + m_t^2 & = & \hat{\alpha}^2 + \hat{\beta}^2 +
  \hat{\gamma}^2 + \hat{\kappa}_2^2 + \hat{\kappa}_3^2
\end{eqnarray}
As a consequence, $U$ is a function of four real parameters (and, of course,
the masses of the up-quarks),
\begin{equation}
\label{uabh}
  U \; = \; U(\hat{\alpha}, \hat{\beta}, \psi_2, \psi_3 ) \; \; \; ,
\end{equation}
see appendix~A for more details.\\
Next we turn to the down-sector. In order to guarantee the NNI form in the
down-sector, too, the unitary matrix $U$, eq.~(\ref{upre}), has to fulfill one
additional condition. With $\widehat{H}^{(d)}=U^\dagger H^{(d)}U$ and setting
$U= (u_{ij})$ and $H^{(d)}=V_{\rm CKM} D^{(d)} V_{\rm CKM}^\dagger =(h_{ij})$, see
(\ref{startpoint}), this condition reads
\begin{equation}
\label{nnidown}
  ( \widehat{H}^{(d)} )_{12} \; = \; u_{i1}^\star h_{ij} u_{j2} 
  \; \stackrel{!}{=} \; 0 \; \; \; .
\end{equation}
This yields the second condition for the matrix $U$.  As this is an equation for
complex numbers, two out of the four parameters $\hat{\alpha},
\hat{\beta}, \psi_2$ and $\psi_3$ are fixed this way, leaving a freedom of
two real parameters.  This is characteristic for the NNI form. Please note that
it is essential to take into account properly the freedom parametrized in $P$,
eq.~ (\ref{phase}).  Had we missed the phase matrix $P$, $U$ would have been
completely determined by (\ref{nnidown}) and only \emph{one} special set of NNI
mass matrices would have resulted, contrary
to our purpose of reconstructing \emph{all} NNI mass matrices.\\
The parameterization of $U$, eq.~(\ref{upre}), in terms of $\hat{\alpha},
\hat{\beta}, \psi_2$ and $\psi_3$ is not suited for constructing the
solutions of (\ref{nnidown}), simply because the resulting equation contains a
complicated sum of cosines with arguments $\psi_2$, $\psi_3$, as well as their
difference $\psi_2 - \psi_3$. The practical reconstruction in this framework
would not be simpler than within the original proposal of Branco, Lavoura and
Mota.  The situation changes decisively if we use a new parameterization of $U$
in terms of two complex numbers defined as follows\footnote{The case $u_{12}=0$
  must be excluded at this point. However, as we shall see below, this is no
  restriction: The case where $a$ tends to infinity, $a\to\infty$, is mapped, by
  a symmetry of the equations, to the point $a=0$, cf. eq.~(\ref{subs}) below.}
\begin{equation}
\label{newvar}
  a := \displaystyle\frac{u_{22}}{u_{12}} \; \; \; \mbox{ and } \; \; \;
  b := \displaystyle\frac{u_{32}}{u_{12}} \; \; \; .
\end{equation}
In particular, the moduli and phases of $a$ and $b$ are given by
\begin{eqnarray}
  \label{mod_a}
  |a| & = & \sqrt{
  \frac{(\hat{\beta}^2-m_u^2)(m_c^2-\hat{\alpha}^2)
        (m_t^2-m_u^2)}
       {(\hat{\alpha}^2-m_u^2)(\hat{\beta}^2-m_c^2)
        (m_t^2-m_c^2)}} \, , \\
  \label{mod_b}
  |b| & = & \sqrt{
  \frac{(\hat{\beta}^2-m_u^2)(m_t^2-\hat{\alpha}^2)
        (m_c^2-m_u^2)}
       {(\hat{\alpha}^2-m_u^2)(m_t^2-\hat{\beta}^2)
        (m_t^2-m_c^2)}} \, ,
\end{eqnarray}
\begin{eqnarray}
  \label{phase_a}
  \psi_a & = & \left\{ 
    \begin{array}{ccc}
\psi_2 & \mbox{for} & m_u^2\le\hat{\alpha}^2\le
m_c^2\le\hat{\beta}^2 \le m_t^2 \\
\psi_2+\pi & \mbox{for} & m_u^2\le\hat{\beta}^2\le
m_c^2\le\hat{\alpha}^2 \le m_t^2
    \end{array}\right. \, , \\
  \label{phase_b}
  \psi_b & = & \psi_3 \, . 
\end{eqnarray}
The two complex variables $a$ and $b$ replace the four real variables
$\hat{\alpha}, \hat{\beta}, \psi_2$ and $\psi_3$. In fact, a
straightforward calculation shows that $U$, eq.~(\ref{upre}), when expressed in
terms of the new variables $a$ and $b$ reads as follows:
\begin{equation}
\label{u}
  U  =  \left( \begin{array}{ccc}
  \displaystyle\frac{(m_t^2 - m_c^2) |a| |b|}{N_2} & 
  \displaystyle\frac{1}{N_1} &
  - \displaystyle\frac{|a|^2 (m_c^2 - m_u^2) + 
                       |b|^2 (m_t^2 - m_u^2)}{N_1 N_2} \\[2ex]
  - \displaystyle\frac{(m_t^2 - m_u^2) |b| a}{N_2 |a|} &
  \displaystyle\frac{a}{N_1} &
  - \displaystyle\frac{a (|b|^2 (m_t^2 - m_c^2) 
                          - (m_c^2 - m_u^2))}{N_1 N_2} \\[2ex]
  \displaystyle\frac{(m_c^2 - m_u^2) |a| b}{N_2 |b|} &
  \displaystyle\frac{b}{N_1} &
  \displaystyle\frac{b (|a|^2 (m_t^2 - m_c^2)
                        + (m_t^2 - m_u^2))}{N_1 N_2}
  \end{array} \right) 
\end{equation}
\begin{eqnarray}
  \label{unorm}
  \mbox{with } N_1^2 & = & 1 + |a|^2 + |b|^2 \\
  \mbox{and } N_2^2 & = & |ab|^2 (m_t^2 - m_c^2)^2 +
  |a|^2 (m_c^2 - m_u^2)^2 + |b|^2 (m_t^2 - m_u^2)^2 \nonumber
\end{eqnarray}
As long as only the up-sector is under consideration $a$ and $b$ remain
arbitrary and are not restricted at all. It is the second NNI condition
(\ref{nnidown}) that imposes a constraint on them: in the new parameterization
this condition takes the simple form of a \emph{quadratic} equation, viz.
\begin{eqnarray}
\label{constr}
  (m_t^2 - m_c^2) \; a b \; (h_{11} + a h_{12} + b h_{13}) & & \nonumber \\
  - \; (m_t^2 - m_u^2) \; b \; (h_{21} + a h_{22} + b h_{23}) & & \nonumber \\
  + \; (m_c^2 - m_u^2) \; a \; (h_{31} + a h_{32} + b h_{33}) & = & 0
\end{eqnarray}
Depending on whether (\ref{constr}) is solved for $a = a(b)$ or for $b = b(a)$
the complex parameter $b$ or the complex parameter $a$ remains free and, thus,
we recover the two-parameter freedom of the NNI
reconstruction.\\
By means of the above formulae elementary calculations yield all parameters of
the NNI mass matrices in terms of $a$ and $b$. For instance, for
$\hat{\beta}$ we obtain:
\begin{equation}
\label{loesbeta}
  \hat{\beta}^2 \; = \; \frac{m_u^2 + |a|^2 m_c^2 + |b|^2 m_t^2}
                             {1 + |a|^2 + |b|^2}
\end{equation}
After insertion of $a = a(b)$ or $b = b(a)$ according to (\ref{constr}) this
equation specifies \emph{all} admissible values for the parameter $\hat{\beta}$
in the NNI form of the mass matrices by varying the unconstrained parameter
$b$ or $a$, respectively.\\
The results for all other parameters are quoted in appendix~B.
\section{Parameter dependencies, symmetries and expansions}
The results derived in the previous section whose details are spelled
out in appendix~B, are completely general and analytical in the sense
that no approximations whatsoever have been made. Furthermore, we
conjecture that the parameterization and reduction to the complex
parameter $a$ (or, alternatively, the parameter $b$) described in the
previous section is the best one can do in reconstructing mass
matrices from their eigenvalues and the CKM observables. In this
section we provide further support for this conjecture by giving some
examples and by showing that it is possible to classify, in a
procedure that is suitable for practical studies, the set of all mass
matrices that are compatible with the given observables.

Generally speaking, due to the use of a (as we conjecture) optimal
parameterization the task of finding \emph{all} mass matrices in NNI form is
reduced to the simple problem of solving a \emph{quadratic} equation, see
(\ref{constr}).  To begin with we note that the left-hand side of
eq.~(\ref{constr}) can be written as a scalar product
\begin{eqnarray}\label{constr1}
  \lefteqn{\left(
    \begin{array}{ccc}
(m_t^2-m_c^2)ab\, , & -(m_t^2-m_u^2)b\, , & (m_c^2-m_u^2)a
    \end{array}\right) \Bigl( h_{ij}\Bigr)\left(
    \begin{array}{c}
1 \\ a \\ b
    \end{array}\right)} \nonumber \\
 & = &  (m_t^2-m_c^2)ab \;\left(
\begin{array}{ccc}
1\, , & -\mu_1/a\, , & \mu_2/b
\end{array}\right) \Bigl( h_{ij}\Bigr) \left(
    \begin{array}{c}
1 \\ a \\ b
    \end{array}\right) = 0\, ,
\end{eqnarray}
where the constants $\mu_1$ and $\mu_2$ denote the ratios
\begin{equation}
  \label{ratios}
  \mu_1 = \frac{m_t^2-m_u^2}{m_t^2-m_c^2}\, ,\quad 
  \mu_2 = \frac{m_c^2-m_u^2}{m_t^2-m_c^2}=\mu_1 - 1\, .
\end{equation}
When the condition (\ref{constr1}) is written in this form and using the fact
that $\widehat{H}^{(d)}$ is hermitean, we see at once that if $(a,b)$ is a
solution, so is $(-\mu_1/a^*,\mu_2/b^*)$. The simultaneous substitution
\begin{equation}
  \label{subs}
  \left(
    \begin{array}{c}
a \\ b
    \end{array}\right) \longmapsto \left(
  \begin{array}{c}
a^\prime = -\mu_1/a^* \\
b^\prime =  +\mu_2/b^*
  \end{array}\right)
\end{equation}
maps the circle with radius $R_a=\sqrt{\mu_1}$ in the complex $a$-plane onto
itself (by relating antipodes), while in the complex $b$-plane every point of
the circle $\sqrt{\mu_2}e^{\I\phi_b}$ is a fixed point. At the same time, this
substitution means interchanging the first and second columns of $U$,
eq.~(\ref{u}). Therefore, if we restrict, e.g., $a=a_{\rm inner}$ to the
interior of the first circle, and calculate $b(a_{\rm inner})$ from
(\ref{constr}) as well as the mass matrices $\bigl(
\widehat{H}^{(u),(d)}\bigr)_{\rm inner}$, then the solution pertaining to
$a_{\rm outer}\equiv a^\prime=-\mu_1/a_{\rm inner}^*$ and the corresponding
value of $b^\prime (a_{\rm outer})$ yields the mass matrices
\begin{displaymath}
  \bigl( \widehat{H}^{(u),(d)}\bigr)_{\rm outer} = U_0^\dagger
  \bigl( \widehat{H}^{(u),(d)}\bigr)_{\rm inner} U_0\quad \mbox{with}
\end{displaymath}
\begin{displaymath}
  U_0 = \left(
    \begin{array}{ccc}
0 & 1 & 0 \\ 1 & 0 & 0 \\ 0 & 0 & 1
    \end{array}\right)\, .
\end{displaymath}
In particular, for unprimed and primed parameters of the triangular matrices
this is equivalent to
\begin{displaymath}
  \bigl( \alpha^2\bigr)_{\rm inner} = \bigl( \beta^2\bigr)_{\rm outer} \, \quad
  \bigl(\beta\kappa_2\bigr)_{\rm inner} = \bigl(\alpha\kappa_3\bigr)_{\rm
    outer}\, ,
\end{displaymath}
\begin{displaymath}
  \bigl( \gamma^2 + \kappa_2^2 + \kappa_3^2\bigr)_{\rm inner} =
  \bigl( \gamma^2 + \kappa_2^2 + \kappa_3^2\bigr)_{\rm outer}\, ,
\end{displaymath}
and likewise for the primed quantities: 
\begin{displaymath}
  \bigl( \alpha^{\prime\, 2}\bigr)_{\rm inner} = 
  \bigl( \beta^{\prime\, 2}\bigr)_{\rm outer} \, \quad
  \bigl(\beta^\prime\kappa_2^{\prime}e^{\I\varphi_2^\prime}
  \bigr)_{\rm inner} = 
  \bigl(\alpha^\prime\kappa^{\prime}_3e^{\I\varphi_3^\prime}
  \bigr)_{\rm outer}\, ,
\end{displaymath}
\begin{displaymath}
  \bigl( \gamma^{\prime\, 2} + \kappa_2^{\prime\, 2} + \kappa_3^{\prime\,
    2}\bigr)_{\rm inner} = 
  \bigl( \gamma^{\prime\, 2} + \kappa_2^{\prime\, 2} + \kappa_3^{\prime\,
    2}\bigr)_{\rm outer}\, . 
\end{displaymath}
Clearly, this symmetry simplifies greatly any practical analysis.

Before turning to the general case we illustrate our formulae by a few
special cases:\\ 
(i) If the parameter $a$ vanishes, $a=0$, eq.~(\ref{constr}) gives
$b(a=0)=-h_{21}/h_{23}$. For the $u$-sector we then obtain
\begin{displaymath}
  \hat{\alpha}^2 = m_c^2\, ,\quad \hat{\beta}^2 = 
  \frac{m_u^2|h_{23}|^2+m_t^2|h_{21}|^2}
       {|h_{23}|^2+|h_{21}|^2}\, ,
\end{displaymath}
\begin{displaymath}
  \hat{\kappa}_3 = 0\, ,\quad
  \hat{\beta}\hat{\kappa}_2 = 
  \frac{|h_{21}||h_{23}|(m_t^2-m_u^2)}
       {|h_{23}|^2+|h_{21}|^2}\, ,
\end{displaymath}
\begin{displaymath}
  \hat{\gamma}^2 = \frac{m_u^2m_t^2}{\hat{\beta}^2}\, .
\end{displaymath}
Similarly, for the $d$-sector we find in this case
\begin{displaymath}
  \hat{\alpha}^{\prime\, 2}= h_{22}\, ,\quad 
  \hat{\beta}^{\prime\, 2}=\frac{h_{11}|h_{23}|^2+h_{33}|h_{21}|^2 
  - 2\R\bigl( h_{12}h_{13}h_{23}\bigr)}
  {|h_{23}|^2+|h_{21}|^2}\, ,
\end{displaymath}
\begin{displaymath}
  \hat{\alpha}^\prime\hat{\kappa}_3^\prime
  e^{\I\hat{\varphi}_3^\prime} = 
  \sqrt{|h_{23}|^2+|h_{21}|^2}\frac{h_{12}}{|h_{12}|}e^{\I\psi_2}\, ,
\end{displaymath}
\begin{displaymath}
  \hat{\beta}^\prime\hat{\kappa}_2^\prime
  e^{\I\hat{\varphi}_2^\prime} = 
  \frac{|h_{21}||h_{23}|}{|h_{23}|^2+|h_{21}|^2}
  \Bigl( h_{33}-h_{11}+\frac{h_{21}h_{13}}{h_{23}} -
  \frac{h_{23}h_{31}}{h_{21}}\Bigr)\, ,
\end{displaymath}
\begin{displaymath}
  \hat{\gamma}^{\prime\, 2} = 
  \frac{m_d^2m_s^2m_b^2}{\hat{\alpha}^{\prime\, 2}
  \hat{\beta}^{\prime\, 2}}\, .
\end{displaymath}
This example illustrates the power and the simplicity of the
reconstruction procedure: Given the experimental data
(\ref{startpoint}), with $V_{\rm CKM}$ given in an arbitrary, but
fixed parameterization, the above formulae yield all entries of the
triangular matrices (\ref{trinni}), hence the mass matrices of the
$u$- and $d$-sectors in an NNI-basis.

\noindent
(ii) If we set $b=0$, hence $a(b=0)=-h_{31}/h_{32}$ the mass matrix
in the $u$-sector is given by
\begin{displaymath}
  \hat{\alpha}^2 = m_t^2\, ,\quad \hat{\beta}^2 = 
  \frac{m_u^2|h_{32}|^2+m_c^2|h_{31}|^2}
       {|h_{32}|^2+|h_{31}|^2}\, ,
\end{displaymath}
\begin{displaymath}
  \hat{\kappa}_3 = 0\, ,\quad
  \hat{\beta}\hat{\kappa}_2 = 
  \frac{|h_{31}||h_{32}|(m_c^2-m_u^2)}
       {|h_{32}|^2+|h_{31}|^2}\, ,
\end{displaymath}
\begin{displaymath}
  \hat{\gamma}^2 = \frac{m_u^2m_c^2}{\hat{\beta}^2}\, .
\end{displaymath}
For the $d$-sector we obtain the following expressions
\begin{displaymath}
  \hat{\alpha}^{\prime\, 2}= h_{33}\, ,\quad 
  \hat{\beta}^{\prime\, 2}=\frac{h_{11}|h_{32}|^2+h_{22}|h_{31}|^2 
  - h_{32}h_{13}h_{21} - h_{31}h_{23}h_{12} }
  {|h_{32}|^2+|h_{31}|^2}\, ,
\end{displaymath}
\begin{displaymath}
  \hat{\alpha}^\prime\hat{\kappa}_3^\prime
  e^{\I\hat{\varphi}_3^\prime} = 
  -\sqrt{|h_{32}|^2+|h_{31}|^2}\frac{h_{13}}{|h_{31}|}e^{\I\psi_3}\, ,
\end{displaymath}
\begin{displaymath}
  \hat{\beta}^\prime\hat{\kappa}_2^\prime
  e^{\I\hat{\varphi}_2^\prime} = 
  \frac{|h_{32}||h_{31}|}{|h_{32}|^2+|h_{31}|^2}
  \Bigl( h_{22}-h_{11}+\frac{h_{31}h_{12}}{h_{32}} -
  \frac{h_{32}h_{21}}{h_{31}}\Bigr)\, ,
\end{displaymath}
\begin{displaymath}
  \hat{\gamma}^{\prime\, 2} =
  \frac{m_d^2m_s^2m_b^2}{\hat{\alpha}^{\prime\, 2} 
  \hat{\beta}^{\prime\, 2}}\, .
\end{displaymath}
As in the previous example this shows that it is possible to
reconstruct the triangular matrices (\ref{trinni}) from the data and,
from there, the squared mass matrices
$\widehat{H}^{(q)}=\widehat{T}^{(q)}\widehat{T}^{(q)\dagger}$. 

The two preceding examples are degenerate cases because, by setting $a$ (or $b$)
equal to zero, hence fixing $b(0)$ (or $a(0)$, respectively), the remaining
two-parameter freedom is partly ``frozen''. The only freedom left over is
contained in the phases $\psi_2$ or $\psi_3$, respectively, which come from the
phase matrix (\ref{phase}).  Also, the substitution (\ref{subs}) shows that two
more special cases can be obtained where one of the parameters is sent to
infinity. We also remark in passing that, although unrealistic in the light of
the data, one can easily study the even more degenerate case of $a$ and $b$ both
going to zero, for instance via
\begin{displaymath}
  a\to 0\, ,\quad b=\frac{(m_c^2-m_u^2)h_{31}}{(m_t^2-m_u^2)h_{21}}a
  \to 0\, .
\end{displaymath}
In the procedure proposed by Branco et al., this limit corresponds to
the case $\kappa =0$ in (\ref{nnieigen}). While their analysis needs
more care in this case, ours can be extrapolated smoothly to
$(a=0,b=0)$ in the way described above. Thus, there is no obstruction
against choosing $a$ (or $b$) anywhere in the complex plane.

We now turn to the general case but keep in mind the actual values of
the observables (quark masses and CKM angles). We first notice that
with
\begin{equation}
  \label{up_masses}
  \bigl( m_u=5.1\mbox{ MeV},\, m_c=1350\mbox{ MeV},\, m_t=330000\mbox{
    MeV}\bigr)\, , 
\end{equation}
\begin{equation}\label{down_masses}
  \bigl( m_d=8.9\mbox{ MeV},\, m_s=175\mbox{ MeV},\, m_b=5600\mbox{
    MeV}\bigr) 
\end{equation}
the first ratio (\ref{ratios}) is approximately 1 while the second is
very small,
\begin{displaymath}
  \mu_1\approx 1-1.67\times 10^{-5}\, ,\quad
  \mu_2= \mu_1 - 1 \approx 1.67\times 10^{-5}\, .
\end{displaymath}
It seems appropriate to expand our formulae in terms of
$\mu_2=\mu_1-1$. So, for a given value of $a$, the two solutions
$b_{1/2}(a)$ of the quadratic equation (\ref{constr}) are given
by
\begin{eqnarray}\label{b1_approx}
  b_1(a) & = & \frac{h_{31}+a h_{32}}{h_{21} +
      a(h_{22}-h_{11})-a^2h_{12}} a\,\mu_2 + {\cal O}(\mu_2^2)\, ,\\
  \label{b2_approx}
  b_2(a) & = & \frac{h_{21} +
      a(h_{22}-h_{11})-a^2h_{12}}{a h_{13}-h_{23}} + {\cal O}(\mu_2)\, . 
\end{eqnarray}
Whether or not this is a good approximation, in principle, depends on
the range of $a$ and on the matrix elements $h_{ij}$, hence on the
experimental input. In order to estimate its quality it is useful to
compare the product and the sum of the two approximate solutions to
the product and the sum of the \emph{exact} solutions of the quadratic
equation (\ref{constr}). Thus, denoting the above approximations by
$b_i(a)$, the exact solutions by $b_i^{\rm exact}$, we define
\begin{eqnarray}
  \label{deltas}
  \delta_P & := & \frac{b_1(a)b_2(a)}{b_1^{\rm exact}(a) b_2^{\rm
      exact}(a)} -1\, , \nonumber \\
  \delta_S & := & \frac{b_1(a)+b_2(a)}{b_1^{\rm exact}(a) +
    b_2^{\rm exact}(a)} -1 \, .
\end{eqnarray}
If the data are such that $\delta_P$ and $\delta_S$ are small, and
using the fact that the ratio $b_1/b_2$ is proportional to $\mu_2$,
estimates for the approximate solutions are seen to be the following
\begin{displaymath}
  \frac{b_1(a)}{b_1^{\rm exact}(a)} \approx 1 + \delta_P - \delta_S 
  \, ,\quad
  \frac{b_2(a)}{b_2^{\rm exact}(a)} \approx 1 + \delta_S\, .
\end{displaymath}
In practice, i.e. for realistic values of the experimental input, the
quantities (\ref{deltas}), as well as the modulus of the ratio
$b_1/b_2$ are very small. Indeed, with the masses (\ref{up_masses}),
(\ref{down_masses}), and with the following data for the moduli of the
CKM matrix elements \cite{PDG}, (assuming a positive value of the
CP-invariant $\cal J$)
\begin{equation}
  \label{moduli}
  |V_{ud}|=0.9752\, ,\; |V_{us}|=0.2213\, ,\; 
  |V_{cd}|=0.2211\, ,\; |V_{cs}|=0.9744\, ,
\end{equation}
the elements of the hermitean matrix $\widehat{H}^{(d)}=\bigl(
h_{ij}\bigr)\equiv (m_d^2+m_s^2+m_b^2)\bigl( k_{ij}\bigr)$ are found
to be
\begin{displaymath}
  k_{11} = 6.144\times 10^{-5}\, ,\quad k_{22} = 2.584\times 10^{-3}\, ,
  \quad k_{33} = 0.9973\, ,
\end{displaymath}
\begin{displaymath}
  k_{12} = (1.089 + \I\: 2.080)\times 10^{-4}\, ,\;
  k_{13} = 3.352\times 10^{-3} - \I\: 8.4890\times 10^{-6}\, ,\;
\end{displaymath}
\begin{displaymath}
  k_{23} = 4.062\times 10^{-2} + \I\: 6.988\times 10^{-7}\, .
\end{displaymath}
The quantity $\delta_P$ is easily seen to be
\begin{displaymath}
  \delta_P(a) = -\frac{k_{23}}{ak_{13}-k_{23}}\mu_2\, .
\end{displaymath}
As the numerical values of the elements of $\widehat{H}^{(d)}$ are
such that $|k_{23}|/|k_{13}|\approx 12$ this function is regular
over the interior of the circle with radius $R_a=\sqrt{\mu_1}$ and may 
thus be estimated by means of standard techniques of function
theory. We find
\begin{equation}
  \label{delP_est}
  |\delta _P| \le 1.82\times 10^{-5}\, .
\end{equation}
Due to the symmetry (\ref{subs}) this domain is sufficient to cover
all NNI solutions. Estimating $\delta_S(a)$ is a bit more complicated
because, as a function of $a$, it has two poles in that same domain,
very close to each other. Excluding a small circle around these poles
one finds typically
\begin{equation}
  \label{delS_est}
  |\delta _S| \le 0.023\, .
\end{equation}
Also the ratio $b_1/b_2$ as obtained from (\ref{b1_approx}) and
(\ref{b2_approx}) is estimated as follows
\begin{displaymath}
  \left|\frac{b_1}{b_2}\right| \le 0.032\, .
\end{displaymath}
Given the experimental values (\ref{up_masses}), (\ref{down_masses}), and
(\ref{moduli}), $\delta_S$, eq.~(\ref{delS_est}) is the dominant uncertainty.
Thus, in this framework the expressions (\ref{b1_approx}) and (\ref{b2_approx})
are excellent approximations in the interior of the circle with radius $R_a$
except in a small neighbourhood of the two poles of $b_1(a)$,
eq.~(\ref{b1_approx}). Note that the approximations are continuous in the
parameter $a$ which is to say that the reconstructed mass matrices
$\widehat{H}^{(u)}$ and $\widehat{H}^{(d)}$ depend on the remaining freedom in a
continuous manner. This is particularly relevant when studying the dependence of
the mass matrices on the parameter $a$ and when comparing to textures
obtained from specific physical assumptions. 

We illustrate the method by means of two examples in Figs.~1 and 2. These
figures show the parameters $\hat{\alpha}^2$ and $\hat{\beta}^2$ as functions of
the complex parameter $a$ and for the two approximate solutions $b_{1/2}(a)$,
eqs.~(\ref{b1_approx}) and (\ref{b2_approx}), with $a$ chosen from the interior
of the circle with radius $R_a$. The figures show clearly the smallness of the
neighbourhood of the two poles where the approximation breaks down. Note that if
in that region one wishes to use the exact solutions $b_{1/2}^{\rm exact}$ care
must be taken in insuring continuity when the signs of square roots are chosen.\\

\section{Some examples and conclusions}

In a detailed numerical study \cite{Diplom} we have verified that the
procedure that we are proposing, from a practical point of view, is
manageable and transparent, and that all dependencies of the NNI
parameters of the triangular matrices can be illustrated in a simple
manner.

Assumptions about specific textures of the mass matrices obtained on
the basis of some physical conjecture, may or may not be compatible
with the data. The parameterization that we are proposing in this work
is particularly well suited for testing the consistency of any such
model in a simple and transparent manner. We illustrate this statement
by two examples taken from the literature. Suppose \cite{TY} in an NNI
basis the $u$-sector is constrained to be, in addition to the NNI
condition, 
\begin{displaymath}
  M^{(u)}_{\rm model} = \left( 
    \begin{array}{ccc}
0 & r_1 & 0 \\ r_1 & 0 & r_2 \\ 0 & r_2 & r_3 
    \end{array}\right)\, ,\; \mbox{with}\quad 
  r_1,r_2,r_3\;\mbox{ real}\, .
\end{displaymath}
The corresponding hermitean, squared form is
\begin{displaymath}
  H_{\rm model}^{(u)} = \left(
    \begin{array}{ccc}
r_1^2 & 0 & r_1r_2 \\ 0 & r_1^2+r_2^2 & r_2r_3 \\ r_1r_2 & r_2r_3 &
r_2^2+r_3^2 
    \end{array}\right)
\end{displaymath}
with $r_3^2=m_u^2+m_c^2+m_t^2-2(r_1^2+r_2^2)$. Comparing this to
$\widehat{H}^{(u)}=\widehat{T}^{(u)}\widehat{T}^{(u)\dagger}$ we see
that 
\begin{displaymath}
  \hat{\alpha}^2=r_1^2\, ,\; \hat{\beta}^2=r_1^2+r_2^2\, ,\;
  \hat{\alpha}\hat{\kappa}_3=r_1r_2\, ,\;
  \hat{\beta}\hat{\kappa}_2=r_2r_3\, .
\end{displaymath}
Thus, by eqs.~(\ref{mod_a}), (\ref{mod_b}) the moduli of $a$ and $b$
are fixed and the constraint (\ref{constr}) is reduced to an equation
for the pase factors $e^{\I\psi_a}$ and $e^{\I\psi_b}$. It is then
easy to decide whether or not this equation has a solution and,
thereby, whether or not the ansatz of the model is compatible with the 
data.

The second example \cite{TY} makes again use of an NNI basis but now
constrains the $d$-sector further by assuming
\begin{displaymath}
  M^{(d)}_{\rm model} = \left( 
    \begin{array}{ccc}
0 & s_1 & 0 \\ s_1 & 0 & s_2 \\ 0 & s_3 & s_3 
    \end{array}\right)\, ,\; \mbox{with}\quad 
  s_1,s_2,s_3\;\mbox{ real}\, .
\end{displaymath}
Like in the previous example the remaining freedom is reduced to two
phase factors which must obey the constraint (\ref{constr}). As the
latter contains the input data, i.e. quark masses and CKM mixing
angles, it is not clear a priori that the model is admissible. We note
in passing that according to \cite{TY} both models, within the
experimental error bars, can indeed be used to parameterize the
data. This is checked in our framework by confirming that
eq.~(\ref{constr}) has solutions of modulus 1. 

The point we wish to make by quoting these examples is the following: while in
general it is difficult to test the compatibility of a specific model ansatz
with the data (wihin their experimental error bars), the model may always be
transformed to an NNI-basis. By converting it to our general form in terms of the
parameters $a$ and $b$, its test in the light of the data
is reduced to checking the simple quadratic equation (\ref{constr}).\\

In summary, we found a new parameterization of squared mass matrices in terms of
the experimental input (eigenvalues and mixing observables) and one complex
parameter that allows to sample the space of solutions in an analytical and
transparent manner. Indeed, from the input: quark masses, matrix elements
$h_{ij}$ as obtained from the CKM data, eq.~(\ref{startpoint}), and a choice of
$a$ (from which $b(a)$ is obtained via (\ref{constr}), or vice versa), the
equations given in appendix~B directly yield the mass matrices (\ref{trinni}).
Thus, by varying the parameter $a$ over the circle with radius $R_a$ in the
complex plane, and using the symmetry (\ref{subs}) we scan the space of all
admissible mass matrices, up to unobservable changes of basis.

We conjecture that this procedure of reconstructing all mass
matrices which are compatible with the data up to (unobservable) changes of
bases, is optimal. We obtained this result by combining the idea of using
general NNI-bases \cite{BLM} with the polar decomposition theorem that allows to
restrict the general analysis to triangular matrices \cite{HS1}, \cite{HS2}. The
formulae that we obtained are sufficiently simple to handle so that they may be
implemented in a reconstruction routine that also takes account of the
experimental error bars. Alternatively, as demonstrated by the examples we gave,
our method allows for a quick test of compatibility with the data for any
assumed texture in the mass matrices.

Finally, with our knowledge of neutrino oscillations and of the corresponding
mixing matrix increasing, it will eventually be possible to perform the
analogous analysis of the leptonic mass matrices in the standard model.

\appendix

\section{}

This appendix gives some intermediate results which are skipped in the main text
of section~III. We begin with the expressions for $\hat{\kappa}_2$ and
$\hat{\kappa}_3$ in terms of
$\hat{\alpha}$ and $\hat{\beta}$.\\
Inserting $\hat{\kappa}_2^2$ according to (\ref{char3}) into (\ref{char2}) and
making use of (\ref{char1}) leads after a straightforward calculation to
\begin{equation}
\label{kappa3}
  \hat{\kappa}_3^2 \; = \;
  \frac{(\hat{\alpha}^2 - m_u^2)(m_c^2 - \hat{\alpha}^2)
  (m_t^2 - \hat{\alpha}^2)}{\hat{\alpha}^2 (\hat{\beta}^2 - 
  \hat{\alpha}^2)} \; \; \; .
\end{equation}
In a similar way we also get:
\begin{equation}
\label{kappa2}
  \hat{\kappa}_2^2 \; = \;
  \frac{(\hat{\beta}^2 - m_u^2)(\hat{\beta}^2 - m_c^2)
  (m_t^2 - \hat{\beta}^2)}{\hat{\beta}^2 (\hat{\beta}^2 - \hat{\alpha}^2)}
\end{equation}
Defining $U =: P V$, where $U$ is given by (\ref{upre}), and $V = (v_{ij})$
we thus obtain:
\begin{eqnarray}
\label{vpre}
  & v_{i1} = \left( \begin{array}{c}
  - \sqrt{\frac{(\hat{\beta}^2 - m_u^2)(m_c^2 - \hat{\alpha}^2)
                (m_t^2 - \hat{\alpha}^2)}{(m_t^2 - m_u^2)
                (m_c^2 - m_u^2)(\hat{\beta}^2 - \hat{\alpha}^2)}} \\[2ex]
  \mp \sqrt{\frac{(\hat{\alpha}^2 - m_u^2)(\hat{\beta}^2 - m_c^2)
                (m_t^2 - \hat{\alpha}^2)}{(m_t^2 - m_c^2)
                (m_c^2 - m_u^2)(\hat{\beta}^2 - \hat{\alpha}^2)}} \\[2ex]
  + \sqrt{\frac{(\hat{\alpha}^2 - m_u^2)(m_c^2 - \hat{\alpha}^2)
                (m_t^2 - \hat{\beta}^2)}{(m_t^2 - m_c^2)
                (m_t^2 - m_u^2)(\hat{\beta}^2 - \hat{\alpha}^2)}}
  \end{array} \right) \; , \; 
  v_{i2} = \left( \begin{array}{c}
  - \sqrt{\frac{(\hat{\alpha}^2 - m_u^2)(\hat{\beta}^2 - m_c^2)
                (m_t^2 - \hat{\beta}^2)}{(m_t^2 - m_u^2)
                (m_c^2 - m_u^2)(\hat{\beta}^2 - \hat{\alpha}^2)}} \\[2ex]
  \pm \sqrt{\frac{(\hat{\beta}^2 - m_u^2)(m_c^2 - \hat{\alpha}^2)
                  (m_t^2 - \hat{\beta}^2)}{(m_t^2 - m_c^2)
                  (m_c^2 - m_u^2)(\hat{\beta}^2 - \hat{\alpha}^2)}} \\[2ex]
  + \sqrt{\frac{(\hat{\beta}^2 - m_u^2)(\hat{\beta}^2 - m_c^2)
                (m_t^2 - \hat{\alpha}^2)}{(m_t^2 - m_c^2)
                (m_t^2 - m_u^2)(\hat{\beta}^2 - \hat{\alpha}^2)}}
  \end{array} \right) & \nonumber \\[2.5ex]
  & v_{i3} = \left(
  \sqrt{\frac{(\hat{\alpha}^2 - m_u^2)(\hat{\beta}^2 - m_u^2)}{(m_t^2 - m_u^2)
              (m_c^2 - m_u^2)}} \; , \; 
  - \sqrt{\frac{(m_c^2 - \hat{\alpha}^2)
                (\hat{\beta}^2 - m_c^2)}{(m_t^2 - m_c^2)
                (m_c^2 - m_u^2)}} \; , \;
  \sqrt{\frac{(m_t^2 - \hat{\alpha}^2)
              (m_t^2 - \hat{\beta}^2)}{(m_t^2 - m_c^2)
              (m_t^2 - m_u^2)}}
  \right)^T &
\end{eqnarray}
In (\ref{vpre}) the upper signs for $v_{21}$ and $v_{22}$ refer to the case
\begin{equation}
\label{case1}
  m_u^2 \leq \hat{\alpha}^2 \leq m_c^2 \; \; \; \mbox{ and } \; \; \;
  m_c^2 \leq \hat{\beta}^2 \leq m_t^2 \; \; \; \mbox{ (case I)} \; \; \; ,
\end{equation}
whereas the lower signs pertain to the case
\begin{equation}
\label{case2}
  m_c^2 \leq \hat{\alpha}^2 \leq m_t^2 \; \; \; \mbox{ and } \; \; \;
  m_u^2 \leq \hat{\beta}^2 \leq m_c^2 \; \; \; \mbox{ (case II)} \; \; \; .
\end{equation}
In fact, this distinction of two cases easily follows from the positivity
of $n_i^2$, $i = 1,2,3$, see (\ref{upre}). At the same time, this argument
shows that these two cases exhaust all possibilities.

\section{}
In this appendix we quote the results for the parameters of the mass matrices
in NNI form in terms of the complex variables $a$ and $b$ (\ref{newvar}). For
the up-sector the expressions in question are obtained by means of
(\ref{upcond}) and inserting $U$ according to (\ref{u}), or, equivalently,
by using (\ref{newvar}) and (\ref{kappa3}) - (\ref{vpre}):
\begin{eqnarray}
\label{loesup}
  \hat{\alpha}^2 & = & \frac{1}{N_2^2} \left(
  |a|^2 |b|^2 m_u^2 (m_t^2 - m_c^2)^2 +
  |a|^2 m_t^2 (m_c^2 - m_u^2)^2 +
  |b|^2 m_c^2 (m_t^2 - m_u^2)^2 \right)  \\[1ex]
  \hat{\beta}^2 & = & \frac{1}{N_1^2} \left(
  m_u^2 + |a|^2 m_c^2 + |b|^2 m_t^2 \right)  \\[1ex]
  \hat{\beta} \hat{\kappa}_2 & = & \frac{N_2}{N_1^2} \\[1ex]
  \hat{\alpha} \hat{\kappa}_3 & = & \frac{N_1}{N_2^2}
  |a| |b| (m_t^2 - m_c^2) (m_t^2 - m_u^2) (m_c^2 - m_u^2) 
\end{eqnarray}
In addition, $\hat{\gamma}^2$ is determined from (\ref{char1}), viz.
\begin{equation}
\label{loesup1}
  \hat{\gamma}^2 \; = \; \frac{(m_u m_c m_t)^2}{\hat{\alpha}^2
  \hat{\beta}^2}
\end{equation}
The results for the down-sector follow from
\begin{displaymath}
  \hat{H}^{(d)} \; = \; U H^{(d)} U^\dagger \; \; \; ,
\end{displaymath}
where $H^{(d)}$ is given in (\ref{startpoint}). Setting $H^{(d)} = (h_{ij})$
we thus get:
\begin{eqnarray}
\label{loesdo}
  \hat{\alpha}'^2 & = & \frac{1}{N_2^2}\Bigl\{
  (m_t^2-m_c^2)^2|ab|^2 h_{11} + (m_t^2-m_u^2)^2|b|^2h_{22} +
  (m_c^2-m_u^2)^2|a|^2h_{33} \nonumber \\
  & - & 2(m_t^2-m_u^2)(m_t^2-m_c^2) |b|^2\R (ah_{12}) +
  2(m_c^2-m_u^2)(m_t^2-m_c^2) |a|^2\R (bh_{13}) \nonumber \\
  & - &   2(m_c^2-m_u^2)(m_t^2-m_u^2) \R (abh_{23})\Bigr\}\, ,
   \\[1ex]
  \hat{\beta}'^2 & = & \frac{1}{N_1^2} \Bigl\{
  h_{11} + |a|^2 h_{22} + |b|^2 h_{33} + 2\Bigl( \R (a h_{12})  
  + \R (b h_{13}) +  \R (a b h_{23})\Bigr)\Bigr\}  \\[1ex]
  \hat{\beta}' \hat{\kappa}'_2 e^{i \hat{\varphi}'_2} & = &
  \frac{1}{N_1^2N_2}\Bigl\{ -\bigl[ |a|^2(m_c^2-m_u^2)^2 + |b|^2
  (m_t^2-m_u^2)^2\bigr] (h_{11}+ah_{12}+bh_{13}) \nonumber \\
  & - &
  \bigl[|b|^2(m_t^2-m_c^2) -  (m_c^2-m_u^2)\bigr] a^*
  (h_{21}+bh_{22}+bh_{23}) \nonumber \\
  & + & \bigl[|a|^2(m_t^2-m_c^2) + (m_t^2-m_u^2)\bigr]
  b^* (h_{31}+ah_{32}+bh_{33})\Bigr\} \label{intermediate} \\
  & = &
  \frac{a (h_{31} + a h_{32} + b h_{33}) - 
        b (h_{21} + a h_{22} + b h_{23})}{a b (m_t^2 - m_c^2)}
  \frac{N_2}{N_1^2} \label{final} \\[1ex]
  \hat{\alpha}' \hat{\kappa}'_3 e^{i \hat{\varphi}'_3} & = & 
  \frac{1}{N_1N_2^2|ab|} \Bigl\{ \bigl[ |a|^2(m_c^2-m_u^2) + |b|^2
  (m_t^2-m_u^2)\bigr] \bigl[ -|ab|^2(m_t^2-m_c^2)h_{11}\nonumber \\ 
  & + & a|b|^2 (m_t^2-m_u^2)h_{12} - b|a|^2(m_c^2-m_u^2)h_{13}\bigr] 
   \nonumber \\ 
  & + & \bigl[ |b|^2(m_t^2-m_c^2) - (m_c^2-m_u^2)\bigr] 
  \bigl[ -a^*|ab|^2(m_t^2-m_c^2)h_{21} \nonumber \\ 
  & + & |ab|^2 (m_t^2-m_u^2)h_{22} - a^*b|a|^2(m_c^2-m_u^2)h_{23}\bigr]
   \nonumber \\ 
  & + & \bigl[ |a|^2(m_t^2-m_c^2) + (m_t^2-m_u^2)\bigr] 
  \bigl[ b^*|ab|^2(m_t^2-m_c^2)h_{31} \nonumber \\ 
  & - & ab^*|b|^2 (m_t^2-m_u^2)h_{32} + |ab|^2(m_c^2-m_u^2)h_{33}\bigr]
  \Bigr\}\, . 
\end{eqnarray}
\begin{equation}
  \label{gamma_p}
  \hat{\gamma}^{\prime\, 2} = 
  \frac{(m_dm_sm_b)^2}
       {\hat{\alpha}^{\prime\, 2}\hat{\beta}^{\prime\, 2}} \, .
\end{equation}
The second, alternative form (\ref{final}) of (\ref{intermediate}) is obtained
by a few lines of calculations.


%
%
\begin{figure}
  \begin{center}
  \epsfig{file=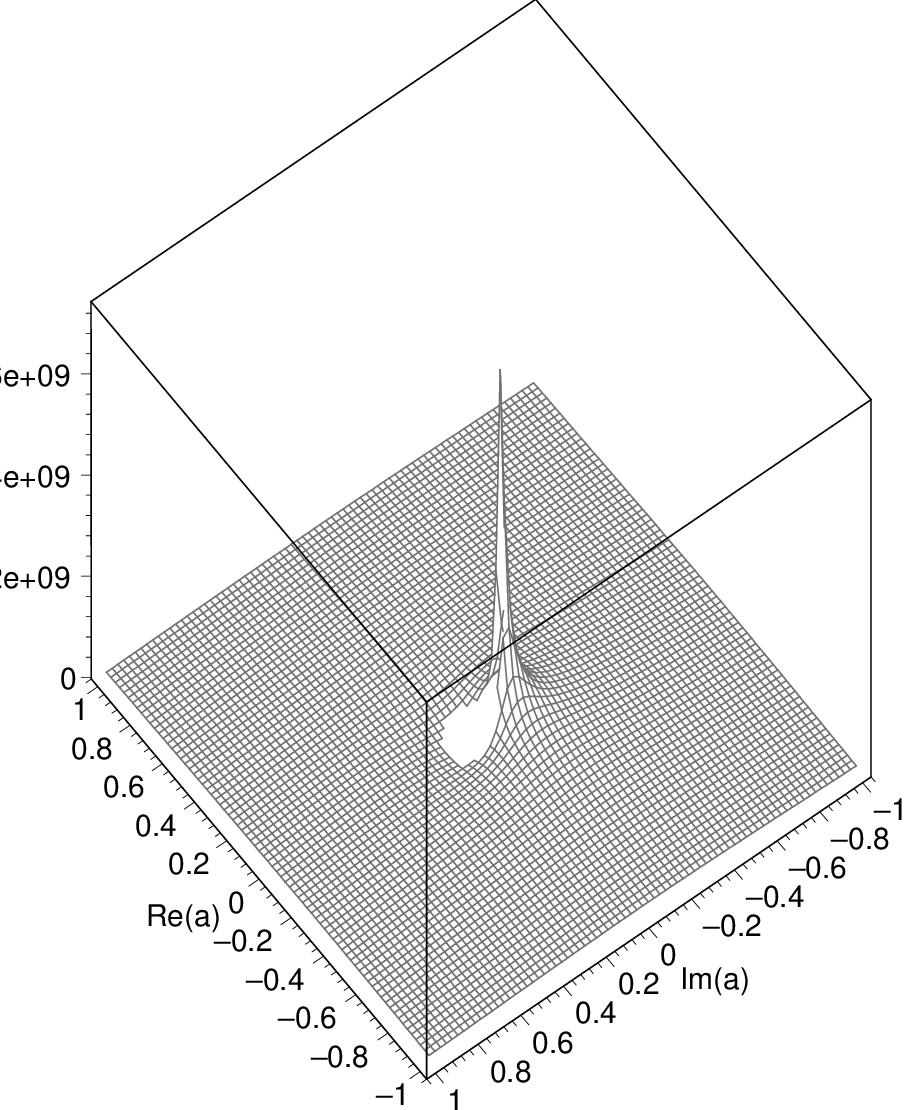,width=7cm}
\vspace{.5cm} 
  \epsfig{file=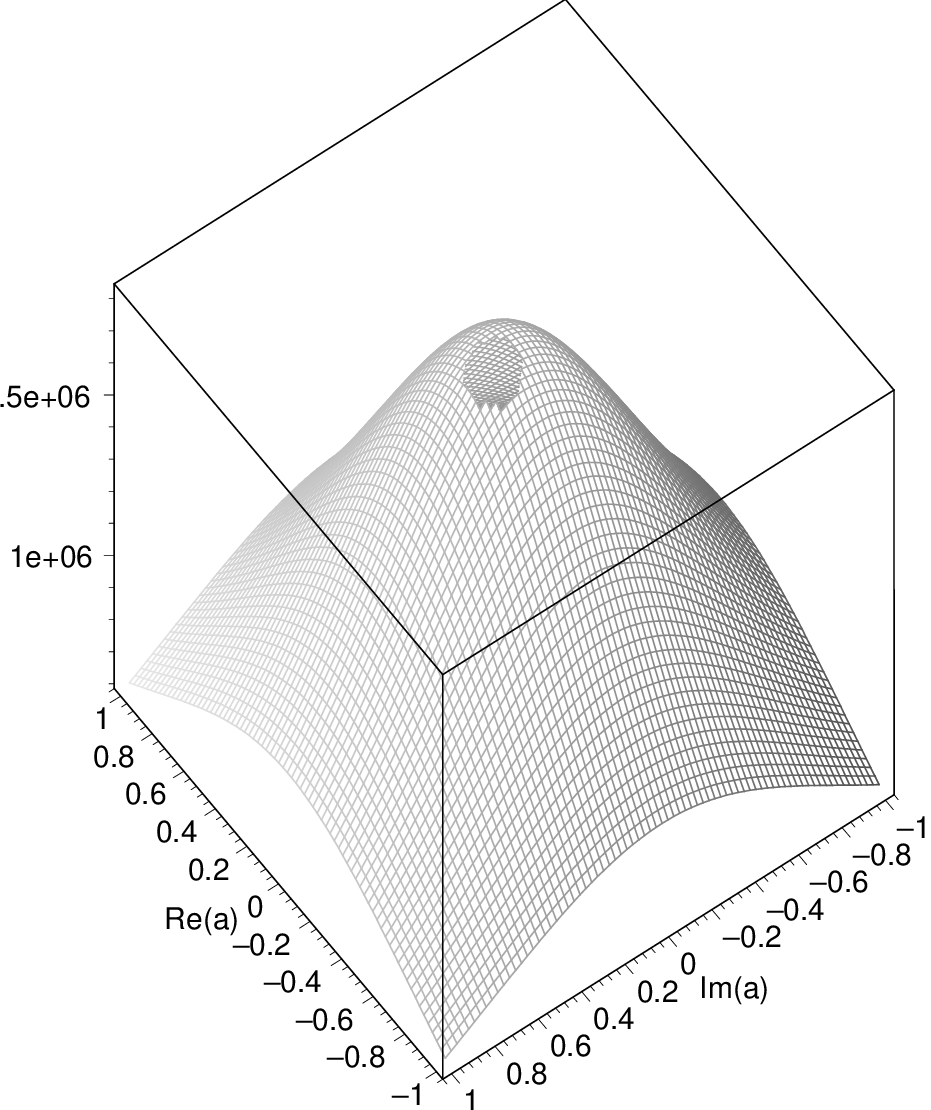,width=7cm}
\caption{Part a: the parameter $\hat{\alpha}^2$ as a function of $a$ (real and
  imaginary part) for the first solution (\ref{b1_approx}); Part b: same
  parameter for the second solution (\ref{b2_approx})}
  \end{center}
\end{figure}
\begin{figure}
  \begin{center}
  \epsfig{file=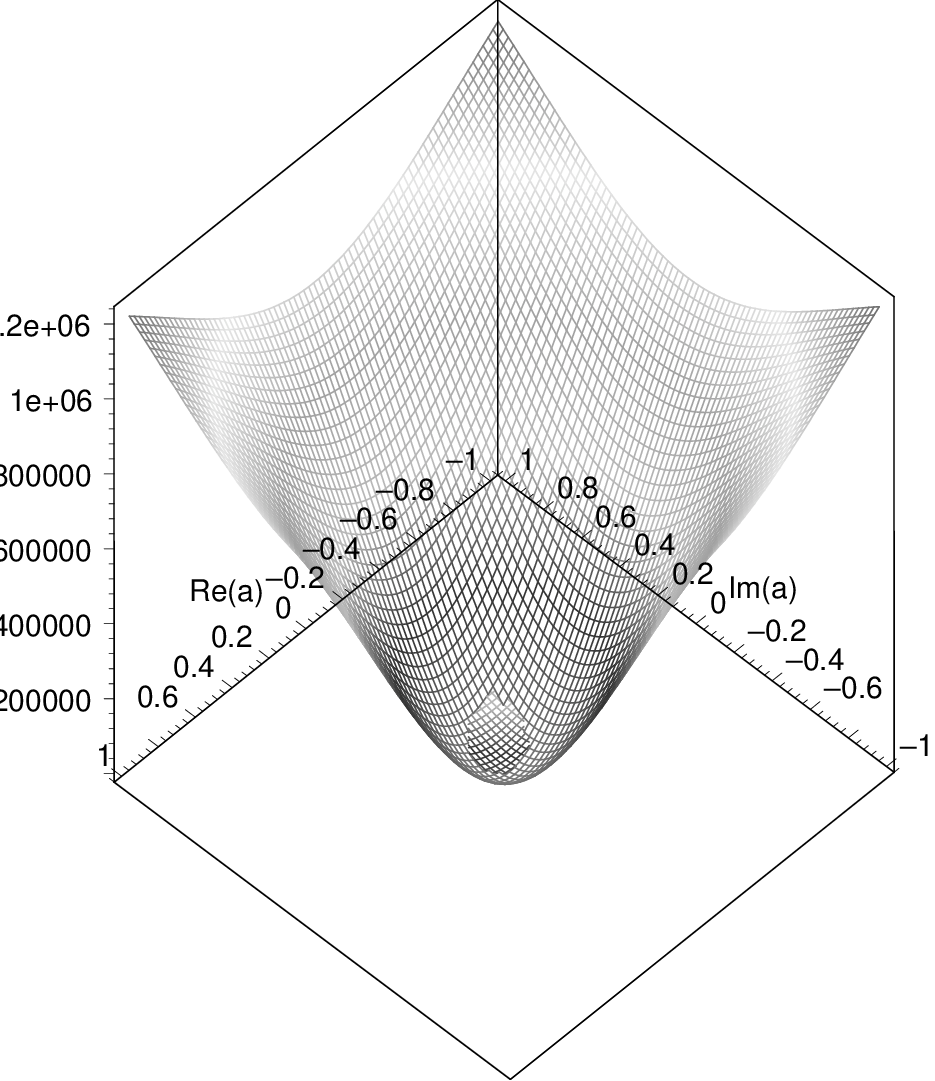,width=7cm}
\vspace{.5cm} 
  \epsfig{file=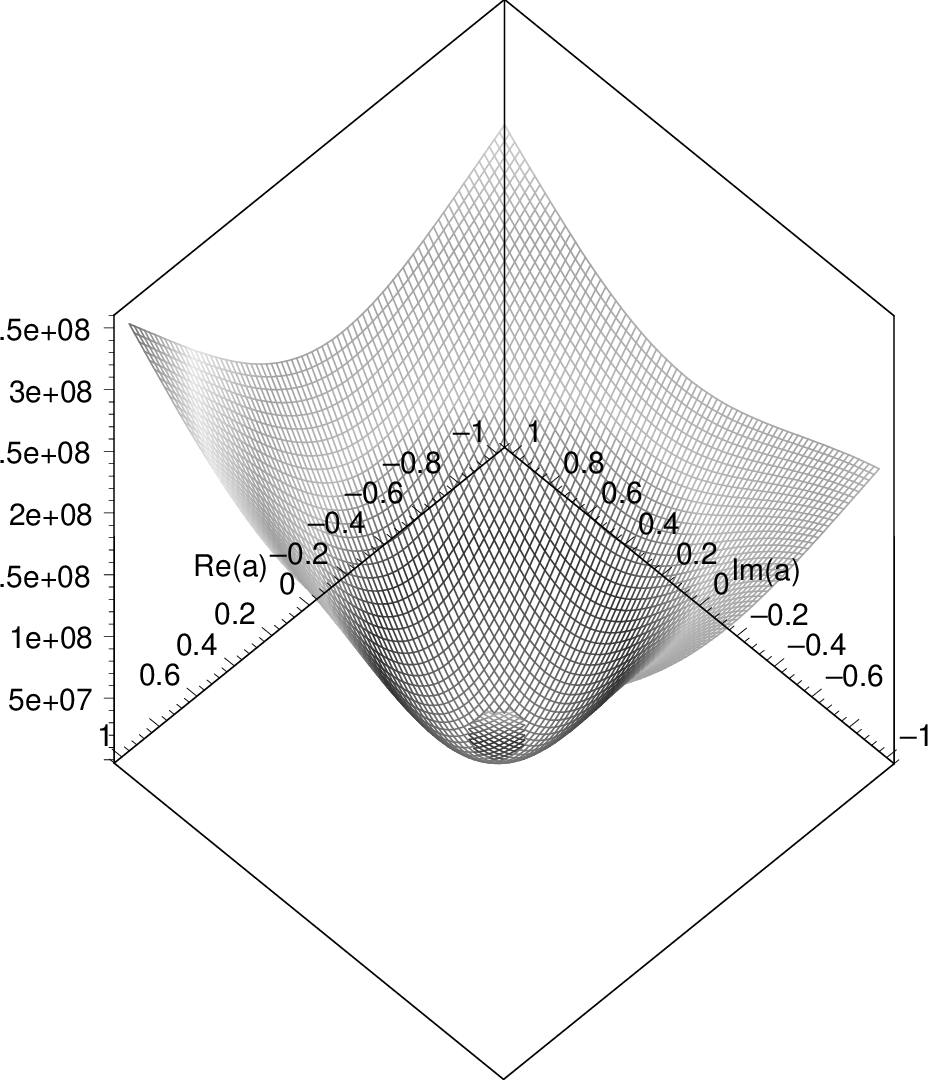,width=7cm}
\caption{Part a: the parameter $\hat{\beta}^2$ as a function of $a$ (real and
  imaginary part) for the first solution (\ref{b1_approx}); Part b: same
  parameter for the second solution (\ref{b2_approx})}
  \end{center}

\end{figure}

\end{document}